\documentclass[twocolumn,tightenlines,showpacs,nofootinbib,preprintnumbers,
amsmath,amssymb]{revtex4-1}

\usepackage{graphicx}
\usepackage{bm}
\usepackage{float}

\begin{document}


\title{Results of measurements of an environment neutron background \\
at BNO INR RAS objects with the helium proportional counter}

\author{V.V.~Alekseenko$^{a}$,
I.R.~Barabanov$^{a}$,
R.A.~Etezov$^{a}$,
Yu.M.~Gavrilyuk$^{a}$, \\
A.M.~Gangapshev$^{a}$,
A.M.~Gezhaev$^{a}$,
V.V.~Kazalov$^{a}$,
A.Kh.~Khokonov$^{b}$, \\
V.V.~Kuzminov$^{a}$,
S.I.~Panasenko$^{c}$,
S.S.~Ratkevich$^{c}$
}
\affiliation{\small $^a$ Institute for Nuclear Research, RAS, Russia \\
$^b$ Kh.M.~Berbekov Kabardino-Balkarian State University, Russia \\
\small $^c$ V.N.~Karazin Kharkiv National University, Ukraine}

\date{\today}

\begin{abstract}
A method of measurements of the environmental neutron background at the Baksan Neutrino Observatory of the INR RAS are described. Measurements were done by using of a proportional counter filled with mixture of Ar(2 at)+$^3$He(4 at). The results obtained at the surface and the underground laboratory of the BNO INR RAS are presented. It is shown that a neutron background in the underground laboratory at the 4900 m w.e. depth is decreased by $\sim 260$ times without any special shield in a comparison with the Earth surface. A neutron flux density in the 5-1323.5~cm air height region is constant within the determination error and equal to $(7.1\pm0.1_{\rm{stat}}\pm0.3_{\rm{syst}})\times10^{-3}$ s$^{-1}\cdot$cm$^{-2}$.

\end{abstract}

\maketitle


\section{\label{intro}Introduction}

A brief description of a new underground laboratory DULB-4900 of the Baksan Neutrino Observatory of the INR RAS is given in the work \cite{DULB}. Experimental search for the Dark Matter WIMP candidates is possible in principle in the DULD-4900 by using of a gas ionization detector with a volume up to 2~m$^3$, solid scintillation detector with a mass up to 300~kg and a double fazes emission detector with a mass of liquid xenon up to 100~kg. Important characteristic which defines sensitivity of such experiments is a background of fast neutrons. Preliminary measurements of fast neutron flux at the place of a future location of the low background rooms of the DULB-4990 
has been done with a test low background set up shielded with the 35~cm thick serpentinite which is a natural mineral with a low natural radioactive elements content \cite{PAN2000Abdurashitov}. It was found that a fast neutron flux with the energy above 700 keV does not exceeds $7\times10^{-8}$~s$^{-1}$cm$^{-2}$ in the shield and is equal to
$(3.5\pm1.7)\times10^{-7}$~s$^{-1}$cm$^{-2}$ without the shield. A 30~l cylindrical volume containing a liquid scintillation (LS) and uniformly distributed nineteen $^3$He proportional counters (PC) was used as a fast neutron detector. The LS served as fast neutrons moderator and the detector of the recoil nuclear energies. The PC registered energy releases of the $^3$He capture reactions with neutrons decelerated up to the thermal energies. This method allows to measure a spectrum of the fast neutrons.
Suitable thermal neutron detector surrounded with a moderator could be used for an estimation only of a fast neutron flux \cite{Thesis_Abdurashitov}. The last method was used for a measurement of neutron flux densities at the low background compartment of the DULB-4900 and others underground and ground-based laboratories of the BNO INR RAS.

\section{Method of measurement}

A detection of thermal neutrons has been done with the CH-04 proportional counter filled
with the [Ar(2~at)+$^3$He(4~at)] gas mixture. A PC box is the stainless steel tube of
30 mm diameter and $\sim0.5$ mm wall thickness. A total PC length
is 980 mm and working length is 900 mm \cite{CH-04}.

A registration of thermal neutrons is due to reaction $^3$He$+n\rightarrow^3$H$+p+(764$~keV)  \cite{Abramov70}. This reaction has a large cross section equal to 5327 b. The calculated CH-04 efficiency is equal to $\varepsilon= 0.68$ \cite{PAN2010Khokonov}.

The measuring equipment included proportional counter, low voltage supply, high voltage supply, charge sensitive preamplifier (CSP) and digital oscilloscope (DO) LAn-20-12PCI incorporated into the personal computer. Positive high voltage has supplied at the anode wire through a filter. Signals from the anode wire entered to the CSP input through a high voltage capacitor.  Amplitudes of pulses from the neutrons were equal to $\sim80$ mV at the $+1400$~V high voltage working value. The signals from the CSP output enters to the DO input. A digitized data record is carried out in the autosynchronizing mode for the pulses with amplitudes exceeding some threshold.

The background pulses from micro discharges and pickups could be presented among the recorded pulses. Usually, such signals have shapes considerably differ from the useful pulse shape. They could be excluded from the analysis by using of the shape discrimination.

\section{Results of measurements}

A number of measurements was carried out during the investigation of the counter
background in different conditions. A neutron background was measured (1) in
a laboratory located at the $2^{\rm{nd}}$ flow of a ground four-storey building, (2) in
the unshielded underground placement DULB-4900, (3) in the shielded
compartment of the DULB-4900 and (4) in the shielded compartment in a
presence of a thermal neutrons source. The measurements (1-2) were made with the PC placed
into the 3~cm cylindrical paraffin moderator and with the uncovered PC.

\begin{figure}[pt]
\includegraphics*[width=2.25in,angle=0.]{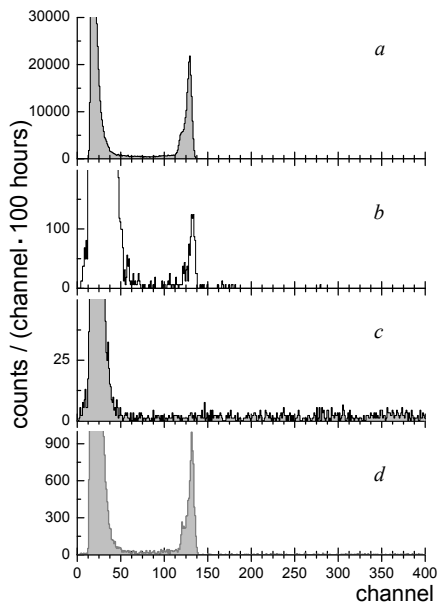}%
\caption{\label{fig:spectra} Amplitude spectra of the uncovered counter. ($a$) -- the 2$^{nd}$ flow of a ground
 four-storey building, ($b$) -- unshielded underground placement DULB-4900, ($c$) -- shielded
 compartment of the DULB-4900,  ($d$) -- [($c$)+thermal neutrons source].}
\end{figure}

The background amplitude spectra were measured with the uncovered PC normalized on 100 hours
measuring time are shown on the Fig.~\ref{fig:spectra}(\emph{a-d}).
The spectra have common features. A peak
corresponding to the 764 keV from the thermal neutrons reactions is seen in the $\sim 130$
channel of the spectra ($a$), ($b$) and ($c$). A shape of the peak for this particular PC is
differs from the usual Gaussian shape by a presence of a step on a left peak slope. A
possible reason of such distortion will be discussed below. A relatively high count rate in
the low energy region ($<300$~keV) could be explained by registration of electrons created
by outer gammas in the counter body and the gas. This is a main source of the background at
the unshielded conditions. The $\beta$-particles appeared in a process of a $\beta$-decay of the impurity natural radioactive $^{40}$K, $^{232}$Th (and daughters), $^{238}$U (and daughters) isotopes in the counter body and of $\beta$-decay of the cosmogenious and man-caused $^{39}$Ar in the counter gas  are another sources of this background. A contemporary activity of the $^{39}$Ar ($T_{1/2}=269$ years)
in the natural argon is equal to $(0.107\pm0.005)$~min$^{-1}\times$(l Ar)$^{-1}$ \cite{Loosli80}.

A flat background component is seen on the spectra at the energies above 300 keV, especially on the spectrum ($c$). It is due to $\alpha$-particles coming to the gas from the inner body surface due to contamination of $^{232}$Th (and daughters) and $^{238}$U (and daughters). A part of the neutron reactions is occurred in the gas near the box wall and only part of the energy release in the gas. Such events give a contribution in the energy region below the neutron peak. A value  \textbf{n}$_f$ of such events depends on a counter geometry and a gas composition and pressure. It is equal to \textbf{n}$_f=0.26$ from a total thermal neutron reactions number for the CH-04 counter. Reactions of the fast neutrons having energies above the thermal one can give a contribution to the spectrum above the neutron peak. An effective cross section of the $^3$He$(n,p)^3$H reaction is inversely proportional to the neutron velocity up to the $\sim100$ keV energy  \cite{Abramov70}.  A total kinetic energy of the reaction products has shifted at the fast neutron kinetic energy in a comparison with the energy release for the thermal neutron. Such event can give a noticeable contribution in the spectrum measured at the unshielded ground conditions.

A measurement with a neutron source was made with purpose to understand a reason of a
difference of the measured neutron peak shape from the Gaussian shape measured usually by
other similar counters. The source was made on a base of $^{238}$Pu (48.2 kBq), $^{239}$Pu (3.58 kBq), $^{226}$Ra (39 kBq) and ($^{233}$U+$^{238}$Pu+$^{239}$Pu) (40.9 kBq) $\alpha$-sources. Their active surfaces were  covered by a 0.2~mm thick beryllium foil. Total neutron flux from this source was estimated as $\sim3$~s$^{-1}$. A box with the
source was surrounded by a 10~cm thickness polyethylene as a
moderator to produce the thermal neutrons. The source was situated above the counter which
was covered by a 1~mm thickness cadmium foil along the length
except the 15~cm central part. The cadmium absorbs effectively
the thermal neutrons and used configuration allows to form a neutron flux in the counter. A
measured spectrum shown on the Fig.~\ref{fig:spectra}($d$). It is seen that a shape of the spectra registered by the 15~cm central counter part 
is similar to the one registered by the counter as a whole. It could be made a conclusion that the shape  distortion could be connected with the longitudinal anode wire defect rather then the end effects. For example, a longitudinal non cylindrical shape of the anode wire could be a reason of such distortion.

Count rates of the CH-04 counter for all data sets are shown on the Table \ref{T1} for the given
energy regions. 
\begin{table*}
\caption{\label{T1}The detector count rate at different conditions.}
\begin{tabular}{c l l r c r c r c r c r c r c c}
\hline \hline
 ~&  ~                         &  ~&\multicolumn{11}{c}{Energy, keV} & ~ & Thermal neutron   \\
                                      \cline{4-14}
~ & \multicolumn{1}{c}{Place,}& ~&\multicolumn{1}{c}{87.3-320}& ~&\multicolumn{1}{c}{320-523}& ~ &\multicolumn{1}{c}{523-669}& ~ &\multicolumn{1}{c}{669-815}& ~ &\multicolumn{1}{c}{815-961}& ~ &\multicolumn{1}{c}{961-1164}& ~ & flux density $\varphi$,       \\
                         \cline{4-4} \cline{6-6} \cline{8-8} \cline{10-10} \cline{12-12} \cline{14-14}
No.&\multicolumn{1}{c}{conditions}& ~& \multicolumn{11}{c} {Count rate, h$^{-1}$}  & ~&  s$^{-1}$cm$^{-2}$   \\
                      \cline{2-2}   \cline{4-14}  \cline{16-16}
1  &Ground building        &    ~& 5140$\pm$21 & ~& 172$\pm$4   & ~& 173$\pm$4    & ~& 1825$\pm$12 & ~& 1.36$\pm$0.34 & ~& 1.27$\pm$0.33 & ~& ($4.67\pm0.03)\times10^{-3}$ \\
2 & -//- + moderator   & ~&\multicolumn{1}{c}{-//-}& ~& 257$\pm$4   & ~& 139$\pm$3    & ~& 2903$\pm$15 & ~& 3.5$\pm$0.5   & ~& 1.90$\pm$0.38 & ~& ($7.43\pm0.04)\times10^{-3}$ \\
3 & DULB-4900          & ~& 4675$\pm$17 & ~&3.7$\pm$0.5  & ~& 1.2$\pm$0.3  & ~& 10.3$\pm$0.8& ~& 0.31$\pm$0.14 & ~&0.50$\pm$0.18  & ~& ($2.5\pm0.2)\times10^{-5}$   \\
4 &  -//- + moderator   & ~& 4228$\pm$13 & ~&3.0$\pm$0.4  & ~& 1.28$\pm$0.23& ~& 16.4$\pm$0.8& ~&0.30$\pm$0.11  & ~& 0.77$\pm$0.18 & ~& ($4.1\pm0.2)\times10^{-5}$   \\
5 & Compartment          & ~& 41.4$\pm$0.2& ~&0.44$\pm$0.07& ~& 0.32$\pm$0.06& ~&0.37$\pm$0.06& ~& 0.52$\pm$0.08 & ~&$0.41\pm0.06$  & ~& $\leq3.8\times10^{-7}$       \\
6 & -//-+\emph{n}-source& ~& 923$\pm$6   & ~&6.8$\pm$0.5  & ~&6.51$\pm$0.53 & ~&79.0$\pm$1.8 & ~&0.26$\pm$0.11  & ~&0.39$\pm$0.13  & ~& ($2.01\pm0.05)\times10^{-4}$ \\
7 & KAPRIZ              & ~& 1210$\pm$4  & ~&0.71$\pm$0.10& ~&0.48$\pm$0.08 & ~& 2.3$\pm$0.2 & ~& 0.32$\pm$0.07 & ~&0.32$\pm$0.07  & ~& ($4.9\pm0.5)\times10^{-6}$   \\
\hline \hline
\end{tabular}
\end{table*}
A count rate under the thermal neutron peak is written in the 669-815 keV
cells. A data of this column could be used for the neutron flux density $\varphi$ determination
according to the expression $\varphi=4R/(S \varepsilon)$ where $R$ - neutron total count rate,
$S=2\pi r l + 2 \pi r^2$ ($r$ - the radius of PC, $l$ - working length of PC) - area of the working counter surface, $\varepsilon$ -  registration efficiency. The $R$ value could be determined by using of a count rate $R_n$ under the neutron peak according to the expression $R=R_n/(1-{\rm{\textbf{n}}}_f)$. A value $\varphi$ could be calculated as $\varphi=9.22\times10^{-3}\times R_n$ by a substitution of
numerical value of the parameters. A background of the counter measured in the shielded
compartment should be subtracted from the analyzed data in a process of a neutron flux
density determination for an unshielded room.  The obtained results are shown in the last
column.
\begin{table*}
\caption{\label{T2}Count rate of the detector at a different height.
}
\begin{tabular}{c l c r c r c r c r c c c c }
\hline \hline
\multicolumn{3}{c} ~       & \multicolumn{9}{c}{Energy, keV} & ~  & ~ \\
                                \cline{4-12}

\multicolumn{3}{c} ~       & 87.3-320 &~&320-523 &~&523-669 &~&669-815 &~&815-960 & ~& Thermal neutron flux density $\varphi$,       \\
                           \cline{4-4}   \cline{6-6} \cline{8-8} \cline{10-10} \cline{12-12}
No. & {Object}& ~&  \multicolumn{9}{c} {Count rate, h$^{-1}$}  & & ~ $\times 10^{-3}$ s$^{-1}$cm$^{-2}$   \\
  \cline{2-2}                             \cline{4-12}    \cline{14-14}
1& 1323.5 cm & ~& 4107$\pm$64 & ~& 244$\pm$16 & ~& 206$\pm$14 & ~&
2515$\pm$50 & ~&$\sim2$ & ~& 6.44$\pm$0.13 \\
2& 1008.5 cm & ~& 3745$\pm$60 & ~& 289$\pm$17 & ~& 274$\pm$16 & ~&
3174$\pm$55 & ~&-//- & ~& 8.13$\pm$0.14 \\
3& 654.5 cm & ~& 3883$\pm$62 & ~&  292$\pm$17 & ~& 272$\pm$16 & ~&
3269$\pm$57 &  ~&-//-& ~ & 8.37$\pm$0.15 \\
4& 332.5 cm & ~& 3558$\pm$60 & ~& 325$\pm$18 & ~& 263$\pm$16 & ~&
3439$\pm$59 & ~&-//-& ~ & 8.81$\pm$0.15 \\
5& 5.0  cm & ~& 4724$\pm$69 & ~& 320$\pm$18 & ~& 250$\pm$16 & ~& 3476$\pm$59 & ~&-//-& ~ & 8.90$\pm$0.15 \\
6& 5.0 cm; far $(\cdot)$ & ~& 4993$\pm$99 & ~& 278$\pm$23 & ~& 231$\pm$21 & ~&
2852$\pm$75 & ~&-//-& ~ & 7.30$\pm$0.19 \\
7& 5.0 cm+mod. & ~& 4965$\pm$99 & ~& 450$\pm$30 & ~& 361$\pm$27 & ~& 4861$\pm$98 & ~&-//-& ~ & 12.45$\pm$0.25
\\ \hline \hline
\end{tabular}
\end{table*}

The $\varphi/4=(6.2\pm0.5)\times10^{-6}$ s$^{-1}$cm$^{-2}$ value  measured in the DULB-4900 could be compared with the value of $(5.4\pm1.3)\times10^{-7}$ s$^{-1}$cm$^{-2}$ measured in the Gran Sasso underground laboratory \cite{Debicki2009}. A noticeable difference of thermal neutron background levels is visible.   At the same time, the DULB-4900 neutron background is similar one measured in the Sanford Underground Research Facility (TCR location) which is $(8.1\pm0.1\pm0.6)\times10^{-6}$ s$^{-1}$cm$^{-2}$  \cite{Best2015}.

It is seen from a comparison of the data in the 146 keV windows "3"-"5" of the spectrum measured in the compartment that the count rates are similar and neutron peak is absent within a statistics. A magnitude of the data in the window "3" was subtracted from the one in the window "4" to obtain a limitation of the neutron count rate for this case. It was found that the thermal neutron flux in the compartment does not exceed a value $3.8\times10^{-7}$~s$^{-1}$cm$^{-2}$ (90\% C.L.).
A conclusion could be made that the thermal neutron flux inside the compartment was decreased by 70 times at least from a comparison of the thermal neutron fluxes in different places of the DULB-4900. This coefficient could be used for the evaluation of a fast neutron flux $\varphi_{fn}$ inside the compartment because of the neutron count rate in this place does not depend on the moderator presence. A fast neutron flux density upper limit was estimated by using the result of the work \cite{PAN2000Abdurashitov} as $(\varphi_{fn}+1.64\sigma)/70$ where $\varphi_{fn}=3.5\times10^{-7}$ and $\sigma=1.7\times10^{-7}$ is a statistical error and was found to be $\leq9.0\times10^{-9}$~ s$^{-1}$cm$^{-2}$ (90\% C.L.).

\begin{figure}[pt]
\includegraphics*[width=2.75in,angle=0.]{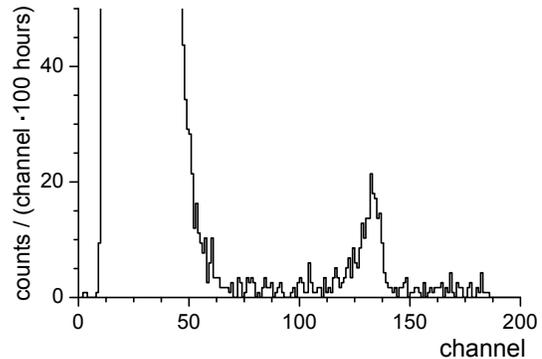}%
\caption{\label{fig:spectra_KAPRIZ} Amplitude spectra of the uncovered counter in the KAPRIZ underground
laboratory.}
\end{figure}

A count rate of the uncovered counter measured in the KAPRIZ underground laboratory located
at the 1000 m w.e. depth is shown in the bottom line of the
Table \ref{T1}. The spectrum is shown on the Fig.~\ref{fig:spectra_KAPRIZ}.
The walls of the KAPRIZ laboratory encased
with a low-radioactivity concrete prepared on a base of the dunite rock. A background of
gammas in the KAPRIZ is decreased at $\sim4$ times in a comparison with the background in the
uncovered cavity or in the unshielded place of the DULB-4900.
A thermal neutron background decreased at 5.2 times.

It is seen from a comparison of the count rates measured in the same place by the uncovered
counter and the counter covered with a moderator that the moderator increased a thermal
neutron flux at 1.6 times at the underground and at the ground surface.  
The moderator efficiently converted epithermal and low energy neutron mainly to the thermal neutron because of a small thickness. 
A correspondence of values of the thermal neutron flux increasing coefficient indicated the similar shapes of a low energy part of the underground and ground neutron spectra.

\begin{figure}[pt]
\includegraphics*[width=3.00in,angle=0.]{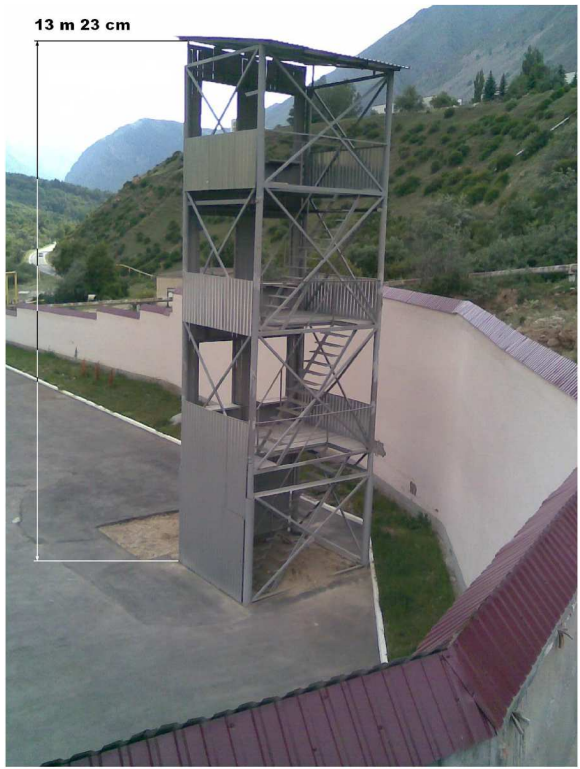}%
\caption{\label{fig:overview} Overview of the fire training tower and the earth location.}
\end{figure}

\begin{figure}[pt]
\includegraphics*[width=3.0in,angle=0.]{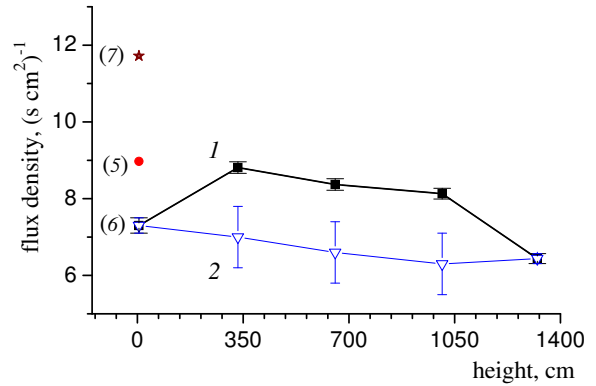}%
\caption{\label{Height_dependence}
Height dependence of a thermal neutron flux density in a ground air layer at the altitude of the BNO INR RAS location.
\emph{1} -- initial data; \emph{2} -- corrected data.}
\end{figure}

The set-up with the CH-04 counter was used to study a thermal neutron flux distribution
above the ground surface at the BNO INR RUS placement altitude ($\sim1700$~m above sea level).
The uncovered counter was placed at different altitudes points of a fire training
tower positioned at an open place sufficiently far from buildings and folds. The
construction materials have relatively small thickness and do not distort a neutron flux
much. An overview of the tower and the earth location are shown on the Fig.~\ref{fig:overview}.
The measurements were done at the 1323.5 cm (point~1), 1008~cm (2), 654.5~cm (3), 332.5~cm (4), 5.0~cm (5) heights.
The point (1) was situated near the roof, the point (5) was on the asphalt and points (2-4) were on the 5~cm thickness wooden
overlaps. An area of a landing is at the bottom of the tower. It prepared by an exchange of a standard soil by a mixture of sand and sawdust in the $\sim 250\times300\times200$~cm$^3$  volume. This place could have an anomaly of the neutron flux density because of a changing of
absorption and moderation behaviors of an upper layer of a soil. An influence of this nonuniformity will decrease with a height
growth due to a geometric factor. A measurement of a surface thermal neutron flux was done on the asphalt point at a 20~m distance from the tower (point 6) to obtain a value of an undistorted flux. A measurement with the counter inserted to the moderator was repeated at
the same place (point~7). A time of a measurement was 1 hour in the point~(1-5) and 0.5 hour in the points (6,7). 
The obtained count rates for different energy range are shown on the Table \ref{T2}. 
The values of a thermal neutron flux corresponding to this count rates are shown in the last column.  
These data form a dependence (1) shown on the Fig.~\ref{Height_dependence}.

It is necessary to take into account that a part of a thermal neutron flux in the points of
(2-4) was created additionally as a result of the neutron moderation in the wood. This
contribution could be estimated on a base of an analysis of the (6) and
(7) data measurements. The tubular paraffin moderator case increased a
thermal neutron flux in the point (6) by
$(5.15\pm0.31)\times10^{-3}$~s$^{-1}$cm$^{-2}$. A contribution of a wooden floor moderator could be estimated as $(0.35\pm0.15_{\rm{syst}})$ of a paraffin one. It gives an absolute value of addition of $(1.8\pm0.11_{\rm{stat}}\pm0.77_{\rm{syst}})\times10^{-3}$~s$^{-1}$cm$^{-2}$. This value should be subtracted from the values of a thermal neutron flux density in the points (2-4). The obtained dependence 2 is shown on the
Fig.~\ref{Height_dependence} also. It is seen that a thermal neutron flux density is constant at the 5.0-1323.5~cm height region within the framework of the
used assumptions. This result is conformed by calculation made in the work \cite{Gromushkin2009}.

\section{Discussion and conclusion}

The neutron flux is decreased at $\sim360$ times at  the deep underground room comparing  with ground surface as 
it is seen from the Table \ref{T2} (point 5) and Table \ref{T1} (object 3). Cosmic rays are the main source of neutrons at a ground surface and a value of a neutron flux could be considered as a constant at the 13~m distance at least above the surface (Fig.~\ref{Height_dependence}, curve 2). A minor ground surface neutron sources are the spontaneous fission of the natural $^{232}$Th, $^{235}$U and $^{238}$U radioactive isotopes ($\sim10$\%
contribution) and the  ($\alpha,n$)-reactions caused by $\alpha$-particles from the decay of the mentioned isotopes and its daughters ($\sim90$\%
contribution) in the reactions with the light elements in the soil \cite{Alekseenko2007}. The first component drops with cosmic rays intensity decreasing. Cosmic ray intensity at the BNO INR RAS location level is equal to $2.0\cdot10^{-2}$~cm$^{-2}\cdot$s$^{-1}$ and on the 4900~m of water equivalent (DULB-4900) is equal to $2.1\cdot10^{-9}$~cm$^{-2}\cdot$s$^{-1}$ \cite{Gavrin91}. The soil thickness decrease the intensity at $\sim10^7$ times. Thus, a net effect of the thermal neutrons created and moderated in the rock was measured in the unshielded point of the DULB-4900.
Flux density of the thermal neutrons created by the natural radioactivity was found to be to $(2.5\pm0.2)\cdot10^{-5}$~cm$^{-2}\cdot$s$^{-1}$.
A parity of the cosmogeneous and radiogenic components is achieved at the $\sim270$~m w.e. depth at a distance of $\sim200$~m from the underground laboratory entrance. It was assume during an interpretation of underground measurements data that the radiogenic component was a constant on a different depth. But, one has to mind that a rate of a neutron creation in a soil depends on soil composition and the radioactive isotopes content. 
A wall concrete surfacing of the underground rooms could change essentially a neutron flux from the wall surface in comparison with the uncoated room. A material composition used for a building of the low background compartments in the DULB-4900 (250~mm polyethylene+1~mm Cd+150~mm Pb) decreases a flux neutron no less than $10^3$ times. The measurement shows that a neutron flux decreased at 70 times at least. A relatively low measured limit of the flux reduction is connected mainly with a level of the own background of the used detector. 
To switch of the neutrons, it is enough to shield a surface of a beryllium with the lavsan film against $\alpha$-particles.

\textbf{Acknowledgments}

The work was carried out in part with the financial support of the Federal Objective Program of the Ministry of Education and Science of the Russian Federation "Research and Development in the 2007-2013 years on the Priority Directions of the Scientific and Technological Complex of the Russia" under the contract No.~16.518.11.7072 and the "Russian Foundation for Basic Research" under the grant No.~14-22-03059.


\end{document}